\documentclass[twocolumn,showpacs,aps,prl]{revtex4-1}
\usepackage{amsmath,amssymb}
\usepackage{graphicx}
\usepackage{epstopdf}
\usepackage[usenames]{color}
\usepackage{latexsym,epsfig}
\begin{document}
\title{Phases, transitions, and boundary conditions in a model of
interacting bosons}

\author{Jamshid Moradi Kurdestany$^1$, Ramesh V. Pai$^2$,
Subroto Mukerjee$^{1,3}$ and Rahul Pandit$^{1,3,}$}
\altaffiliation [ also at]{ Jawaharlal Nehru Centre for Advanced Scientific Research, Jakkur, Bangalore 560 064, India}
\affiliation{$^1$ Department of Physics, Indian Institute of Science, Bangalore 560 012, India}
\affiliation{$^2$ Department of Physics, Goa University, Taleigao Plateau, Goa 403 206, India}
\affiliation{$^3$ Centre for Quantum Information and Quantum Computing, Indian Institute of Science,
Bangalore 560 012, India}

\begin{abstract}

We carry out an extensive study of the phase diagrams of the
extended Bose Hubbard model, with a mean filling of one boson per
site, in one dimension by using the density matrix
renormalization group and show that it contains Superfluid (SF), Mott-insulator
(MI), density-wave (DW) and Haldane-insulator (HI) phases. We show that
the critical exponents and central charges for the HI-DW, MI-HI
and SF-MI transitions are consistent with those for models in the
two-dimensional Ising, Gaussian, and
Berezinskii-Kosterlitz-Thouless (BKT) universality classes,
respectively; and we suggest that the SF-HI transition may be
more exotic than a simple BKT transition. We show explicitly that
different boundary conditions lead to different phase diagrams.

\end{abstract}
\maketitle

Over the last two decades, experiments on cold alkali atoms in
traps have obtained various phases of correlated bosons and
fermions~\cite{reviews}. Microscopic interaction parameters can
be tuned in such experiments, which provide, therefore, excellent
laboratories for studies of quantum phase transitions, such as
those between superfluid (SF) and Mott-insulator (MI) phases in a
system of interacting bosons in an optical
lattice~\cite{fisher89,sheshadri,jaksch}. The SF and MI phases
are, respectively, the protypical examples of gapless and gapped
phases in such systems; in addition, it may be possible to obtain
exotic phases, e.g., density-wave
(DW)~\cite{kuhner98, rvpai05, entspect, kurdestany12}, Haldane-insulator
(HI)~\cite{torre06,berg08}, and supersolid
(SS)~\cite{kovrizhin05} phases, in systems with a dipolar
condensate of $^{52}{\rm Cr}$ atoms~\cite{werner05}. The first
step in developing an understanding of such experiments is to
study lattice models of bosons with long-range
interactions~\cite{goral02}; the simplest model that goes beyond
the Bose-Hubbard model with onsite, repulsive
interactions~\cite{fisher89,sheshadri} is the extended
Bose-Hubbard model (EBHM), which allows for repulsive
interactions between bosons on a site and also on
nearest-neighbor sites~\cite{rvpai05,kurdestany12}.  The
one-dimensional (1D) EBHM is particularly interesting because (a)
quantum fluctuations are strong enough to replace long-range SF
order by a quasi-long-range SF, with a power-law decay of
order-parameter correlations\cite{rvpai05}, and (b) boundary
conditions play a remarkable role here, in so far as they can
modify the bulk phase
diagram~\cite{torre06,berg08,deng12,rossini12}, in a way that is
not normally possible in the thermodynamic
limit~\cite{thermodynamiclimit}.

\begin{figure}
\centering \epsfig{file=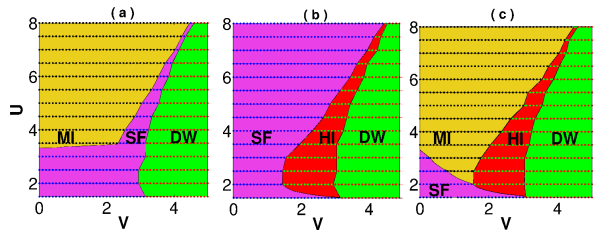,width=8cm,height=3cm}
\caption{\label{fig:PhaseD}{(Color online) Phase diagrams in the $(U,V)$ plane for the
1D EBHM with the following boundary conditions $(a)$ BC1 $(N = L)$,
$(b)$ BC2 $(N = L + 1)$, and $(c)$ BC3 $(N = L)$ and $\mu_r=-\mu_l=2$ showing
Mott-insulator (MI ochre), superfluid (SF purple), Haldane-insulator
(HI red), and density-wave (DW green) phases and the phase boundaries
between them; in this range of $U$ and $V$ all transitions are continuous;
at larger values of $U$ and $V$ the MI-DW and HI-DW transitions become
first-order.}}
\end{figure}

We present extensive density-matrix-renormalization-group
(DMRG)~\cite{rvpai05} studies of the phases, transitions, and the
role of boundary conditions in the 1D EBHM with a mean filling of
one boson per site. Although a few DMRG
studies~\cite{rvpai05,torre06,berg08,deng12,rossini12} have been
carried out earlier, none of them has elucidated completely clearly the
universality classes of the continuous transitions in the 1D
EBHM; nor have they compared in detail, the phase diagrams of
this model with different boundary conditions. Our study, which
has been designed to obtain the universality classes of these
transitions and to examine the role of boundary conditions in stabilizing
different phases, yields a variety of interesting results that we
summarize qualitatively below: If the filling $\rho=1$, in the
thermodynamic limit, we obtain three types of phase diagrams, P1
(Fig.~\ref{fig:PhaseD}(a)), P2 (Fig.~\ref{fig:PhaseD}(b)), and P3
(Fig.~\ref{fig:PhaseD}(c)), for boundary conditions BC1 (number
of bosons $N=L$, where $L$ is the number of sites), BC2
($N=L+1$), and BC3 ($N=L$ and additional chemical potentials at
the boundary sites, as we describe below), respectively. For P1 we
obtain SF, MI, and DW phases and we confirm, as noted
earlier~\cite{rvpai05}, that the
SF-MI transition is in the Berezinskii-Kosterlitz-Thouless
(BKT)~\cite{kogutreview}
universality class and the SF-DW has both BKT and
two-dimensional (2D) Ising characteristics; at sufficiently large values of the
repulsive parameters, the MI-DW transition is of first
order~\cite{rvpai05}. The phase diagram P2 displays SF, HI, and
DW phases, with continuous SF-HI and HI-DW transitions in BKT and
2D Ising universality classes, respectively.
The phase diagram P3 has SF, MI, HI, and DW phases, with SF-MI,
MI-HI, HI-DW in BKT, 2D Gaussian~\cite{gaussian1}, and 2D Ising
universality classes, respectively; the SF-HI transition is
more exotic than a simple BKT transition; at sufficiently large values
of the repulsive parameters, there is a first-order MI-DW phase
boundary.

The 1D EBHM is defined by the Hamiltonian
\begin{equation} {\cal
H} = -t\sum_i (a_{i}^{\dagger} a_{i+1} + h.c)+\frac{U}{2}\sum_i
{\hat n}_i ({\hat n}_i -1)+V\sum_i {\hat n}_i {\hat n}_{i+1},
\label{eq:ebhmodel}
\end{equation}
where $t$ is the amplitude for a boson to hop from site $i$ to
its nearest-neighbor sites, $h.c.$ denotes the Hermitian
conjugate, $a_i^{\dag}, \, a_i$, and ${\hat n}_i \equiv
a_i^{\dag} a_i$ are, respectively, boson creation, annihilation,
and number operators at the site $i$, the repulsive potential
between bosons on the same site is $U$, and $V$ is the repulsive
interaction between bosons on nearest-neighbor sites.
We set the scale of energies by choosing $t = 1$. We work
with a fixed value of the filling $\rho$ and focus on $\rho=1$,
i.e., a filling of one boson per site in the thermodynamic limit,
which we realize in three different ways, to highlight the role
of boundary conditions, by using the three different boundary
conditions BC1, BC2, and BC3.  In all these three cases we have
open boundaries; in BC3, we include boundary chemical potentials
$\mu_l=-2t$ and $\mu_r=2t$ at the left and right boundaries,
which are denoted by the subscripts $l$ and $r$, respectively. We study
this model by using the DMRG method that has been described in
Ref.~\cite{rvpai05}.  To characterize the phases and transitions
in this model, we calculate the following thermodynamic and correlation
functions and the entanglement entropy (familiar from quantum information):
\begin{eqnarray}
&&R_{dw}(\mid i -
j\mid)=(-1)^{\mid i-j\mid}\langle \delta {\hat n}_{i}\delta {\hat
n}_{j} \rangle,\nonumber \\
&&R_{string}(\mid i - j \mid)=\langle
\delta{{\hat n}_{i}} e^{({i\pi \sum_{l=i}^{j}\delta {{\hat
n}_{l}}})} \delta {{\hat n}_j} \rangle,\nonumber \\
&&R_{SF}(\mid i
- j \mid)=\langle a_{i}^{\dagger} a_{j}\rangle, \nonumber \\
&&\mathcal{S}_\pi = \sum_{i,j=1}^{L}e^{i\pi(i-j)}\langle {\hat
n}_{i}{\hat n}_{j} \rangle/L^2,\nonumber \\
&&n(k=0) = \sum_{i,j}^{L}\langle a_{i}^{\dagger} a_{j}\rangle/L,\nonumber \\
&&G_{L}^{NG} = E_{L}^{1}(N)-E_{L}^{0}(N),\nonumber \\
&&G_{L}^{CG} = E_{L}^{0}(N+1)+E_{L}^{0}(N-1)-2E_{L}^{0}(N),\nonumber \\
&&\xi_{L} = \sqrt{\frac{\sum_{i,j=1}^{L}(i-j)^2\langle
a_{i}^{\dagger} a_{j}\rangle} {\sum_{i,j=1}^{L}\langle
a_{i}^{\dagger} a_{j}\rangle}} ,\nonumber \\
&&S_L(l) = -\sum_{i}^{n_{states}}\lambda_i \log_2\lambda_i, \nonumber \\
&&F = LG_{L}^{CG}\left(1+\frac{1}{2\ln L+B}\right), \nonumber\\
&&D = \ln L-a|V-V_{c}|^{-1/2},
\label{eq:measure}
\end{eqnarray}
where  $\delta{\hat n}_i={\hat n}_i-\rho$, the correlation
functions for the DW, HI and SF phases are $R_{dw}(|i-j|)$,
$R_{string}(|i-j|)$, $R_{SF}(|i-j|)$, respectively,
$\mathcal{S}_{\pi}$ is the density-density structure factor at
wave number $k=\pi$, $n(k=0)$ the momentum distribution of the
bosons at $k=0$, $G_{L}^{NG}$ and $G_{L}^{CG}$ are neutral and
charge gaps~\cite{footnote}, $E_{L}^{0}(N)$ and $E_{L}^{1}(N)$
are the ground-state and first-excited-state energies,
respectively, for our system with $L$ sites and $N$ bosons,
$\xi_{L}$ is a system-size-dependent correlation length; $S_L(l)$
is the von-Neumann block entanglement entropy, $\lambda_i$ are
the eigenvalues of the reduced density matrix for the right
block, of length $l$, in our DMRG~\cite{rvpai05}, and
$n_{states}$ is the number of states that we retain for this
density matrix; note that $S_L(l) = c\lambda$ at a critical
point~\cite{kollath}, with $c$ the central charge and
the log-conformal distance
$\lambda=(\log[\frac{2L}{\pi}\sin(\frac{\pi l}{L})])/6$; $a$ and
$B$ depend on the critical value $V_{c}$, at which the concerned
transition occurs; $V_{c}$ depends on $U$. The order parameters
for the DW and HI phases are
$O_{dw}=\sqrt{\lim_{|i-j|\to\infty}R_{dw}(\mid i - j\mid)}$ and
$O_{string}=\sqrt{\lim_{|i-j|\to\infty}R_{string}(\mid i -
j\mid)}$, respectively; for the DW, HI and SF phases, we also use the
Fourier transforms of the correlation functions, which we denote
by $R_{dw}(k)$, $R_{string}(k)$, $R_{SF}(k)$, respectively. Our
calculations have been performed with $100 \leq L \leq 300$, a
maximal number $n_{max}=6$ of bosons at a site, and up to
$n_{states}=256$ states in our density matrices; we
have checked, in representative cases, that, for the ranges of
$U$ and $V$ we cover, our results are not affected significantly
if we use $n_{max}=4$ and $n_{states}=128$.

In Figs.~\ref{fig:PhaseD} (a), (b), and (c) we depict our DMRG
phase diagrams, in the $(U,V)$ plane, for the 1D EBHM with BC1,
BC2, and BC3 boundary conditions, respectively. These phase
diagrams show MI (ochre), SF (purple), HI (red), and DW (green)
phases and the phase boundaries between them; the points inside
these phases indicate the values of $U$ and $V$ for which we have
carried out our DMRG calculations. In the ranges of $U$ and $V$
that we consider here, all transitions are continuous; at larger
values of $U$ and $V$ than those shown in Fig.~\ref{fig:PhaseD},
the MI-DW transitions become first-order, as shown for BC1 in
Ref.~\cite{rvpai05}; the direct MI-DW transition, at large values
of $U$ and $V$, is also of first order with the BC3 boundary
condition. Note that MI (HI) phases appear with boundary
conditions BC1 and BC3 (BC2 and BC3). The MI phase occurs at a
strict commensurate filling (BC1 with $N=L$) and is characterized
by a charge gap, which arises because of the onsite interaction $U$
that penalizes the presence of more than one boson at a site.
With the boundary condition BC2, $N=L+1$, so there is one more
boson in the system than in case BC1; this extra boson can hop
over the commensurate, BC1, MI state, produce gapless excitations
in the upper Mott-Hubbard band, and thereby destroy the MI phase,
which is replaced by SF or HI phases.

The origin of the HI phase in the 1D EBHM can be understood
qualitatively as follows: If $\rho=1$ and $U \gg t$, the MI phase
has one boson at most sites because the large value of $U$
suppresses boson-number fluctuations; as $U$ begins to decrease,
these fluctuations are enhanced and, to leading order in $|t|/U$,
arise from holes at some sites and pairs of bosons at other sites
(occupancies of more than 2 bosons at a site are energetically
disfavored unless $U$ decreases). If we restrict ourselves to
$0,\, 1$, and $2$ bosons at every site, the 1D EBHM can be mapped
onto a spin-$1$ chain, in which the three values of spin
projection correspond to the occupancies $0, 1,$ and $2$. This
spin-$1$ model, an antiferromagnetic chain, with full $SU(2)$
symmetry broken by the extended interaction $V$ in
Eq.(\ref{eq:ebhmodel}), has a Haldane phase, whose analog in the
1D EBHM is the HI phase~\cite{torre06}; this phase also has gapless edge modes,
which destroy the charge gap at strict commensurate filling
(BC1). However, with an extra boson (BC2), these gapless modes get quenched and
so we obtain an HI phase with a gap.

By applying the boundary chemical potentials $\mu_l$ and $\mu_r$
in BC3, we introduce a gap in the edge states in the HI; however,
because the system is exactly at a commensurate filling $N=L$,
the MI phase is also present. Thus, we  obtain both the MI and
the HI phases with the BC3 boundary condition. Note also that the
HI phase exists because of boson-number fluctuations about the
Mott state; so we expect the HI to give way to the MI phase at
large values of $U$ at which such number fluctuations are
suppressed, in much the same way as the SF yields to the MI at
large values of $U$. In the large-$U$ and large-$V$ regions,
which we do not show in the phase diagrams of
Fig.~\ref{fig:PhaseD}, there is a direct, first-order, MI-DW
transition~\cite{rvpai05}.

\begin{figure*} \includegraphics[scale=0.4]{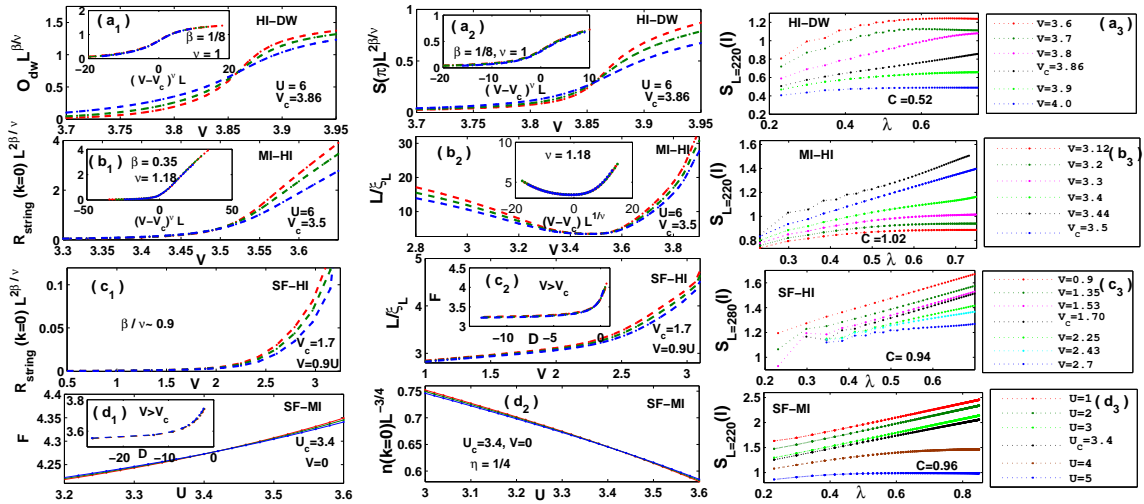}
\caption{\label{fig:OPS}
{(Color online) Plots for the HI-DW transition at $U=6$, with $L=200$ (red
curve), $L=150$ (green curve), and $L=100$ (blue curve), of
$(a_1)$ the scaled DW order parameter $O_{dw}L^{\beta/\nu}$
versus $V$ (the inset shows $O_{dw}L^{\beta/\nu}$ versus
$(V-V_{c})^{\nu}L$), where $\beta$ and $\nu$ are order-parameter
and correlation-length exponents, $(a_2)$ the $k=\pi$ scaled
structure factor $\mathcal{S}{({\pi})}L^{2\beta/\nu}$ versus $V$
(the inset shows $\mathcal{S}{({\pi})}L^{2\beta/\nu}$ versus
$(V-V_{c})^{\nu}L$), and  $(a_3)$ the block entanglement entropy $S_L(l)$
versus the logarithmic conformal distance
$\lambda={\frac{1}{6}}\log[\frac{2L}{\pi}\sin(\frac{\pi l}{L})]$
for an open system of length $L = 220$ for different values of
$V$ (the slope of the linear part of the curve for $V_c=3.86$ yields
a central charge $c = 0.52$). $(b_1)$-$(b_3)$ show the analogs,
for the MI-HI transition, of the plots in $(a_1)$-$(a_3)$ but
with  $(b_1)$ $O_{dw}L^{\beta/\nu}$ replaced by
$R_{string}(k=0)L^{2\beta/\nu}$, the scaled Fourier transform of
string correlation function at $k=0$, and $(b_2)$
$\mathcal{S}{({\pi})}L^{2\beta/\nu}$ replaced by $L/\xi_L$;
in ($b_1$) the values of $L$ are $290$ (red curve) $250$ (green curve),
and $200$ (blue curve) and in ($b_2$) $L$ is $220$ (red curve) $200$
(green curve), and $180$ (blue curve).
$(c_1)$-$(c_3)$ show the analogs, for the SF-HI transition, of
the plots in $(b_1)$-$(b_3)$ (the inset in $(c_2)$ shows a plot
of the scaled charge gap $F$ versus $D$ (see
Eq.~\ref{eq:measure})); in ($c_1$) and ($c_2$) the values of $L$
are $280$ (red curve) $240$ (green curve), and $200$ (blue curve).
$(d_1)$-$(d_3)$ show the analogs, for
the SF-MI transition, of the plots in $(b_1)$-$(b_3)$ but with
$(d_1)$ $R_{string}(k=0)L^{2\beta/\nu}$ replaced by the scaled
charge gap $F$ and $V$ replaced by $U$ and $(d_2)$ $L/\xi_L$
replaced by $n(k=0)L^{-3/4}$; in ($d_1$) $L$ is $200$ (red curve) $180$
(green curve), $160$ (blue curve) and in ($d_2$) $L$ is $250$ (red curve)
$230$ (green curve), and $210$ (blue curve).}}
\end{figure*}

We now obtain the universality classes of the continuous
transitions in the phase diagrams of Fig.~\ref{fig:PhaseD}. It
has been noted in Ref.~\cite{rvpai05} that the MI-SF transition
in Fig.~\ref{fig:PhaseD} (a), with BC1 boundary conditions, is in
the Berezinskii-Kosterlitz-Thouless (BKT) class, whereas the SF-DW
has both BKT and two-dimensional (2D) characters, in as much as
the SF correlation length shows a BKT divergence but the DW order
parameter decays to zero with a 2D-Ising, order-parameter
exponent; this transition should be in the universality class of
the Wess-Zumino-Witten $SU(2)_1$ theory like the critical point
that lies between the BKT phase and the antiferromagnetic phase
in the spin$-1/2$, XXZ antiferromagnetic chain~\cite{shankar90}.
The MI-SF and SF-DW phase boundaries in Fig.~\ref{fig:PhaseD} (a)
merge, most probably at a bicritical point, beyond which the
MI-DW transition is first order~\cite{rvpai05}. We concentrate
here on the HI-SF, HI-DW, HI-MI transitions, all of which can be
found in Fig.~\ref{fig:PhaseD} (c), with the boundary condition
BC3, for which we present results in Fig.~\ref{fig:OPS}; we have
checked explicitly that the universality classes of the HI-SF and
HI-DW transitions are the same with both BC2 and BC3 boundary
conditions.

In Fig.~\ref{fig:OPS} $(a_1)$ we plot the scaled DW order
parameter $O_{dw}L^{\beta/\nu}$ versus $V$ near the
representative point, $V_c \simeq 3.86$ and $U=6$, on the HI-DW
phase boundary for $L=200$ (red curve), $L=150$ (blue curve), and
$L=100$ (green curve); the inset gives a plot of
$O_{dw}L^{\beta/\nu}$ versus $(V-V_{c})^{\nu}L$;
Fig.~\ref{fig:OPS} $(a_2)$ presents a similar plot and inset for
the scaled structure factor $\mathcal{S}{({\pi})}L^{2\beta/\nu}$
at wave number $k=\pi$; in Fig.~\ref{fig:OPS} $(a_3)$ we plot the
von-Neumann block entanglement entropy $S_L(l)$ versus the logarithmic
conformal distance $\lambda$ for $L = 220$ and a range of values
of $V$ in the vicinity of $V_c(U=6)=3.86$. The values of the
exponents $\beta$ and $\nu$ and the value of the central charge
$c$ that we obtain from Figs.~\ref{fig:OPS} $(a_1)-(a_3)$,
namely, $\beta = 1/8, \, \nu = 1$, and $c=0.52$, are consistent
with their values in 2D Ising model. If we restrict the occupancy to one
boson per site, in the limit $U, V \to \infty$ with fixed $U/V$,
the DW state can be represented as $|\dots 20202020 \dots\rangle$
in the site-occupancy basis; this state is doubly degenerate and
clearly breaks the translational symmetry of the lattice by
doubling the unit cell; in contrast, the HI state has the
translational invariance of the lattice. Given this double
degeneracy of the DW state, we expect that, if the HI-DW
transition is continuous, then it should be in the 2D Ising
universality class; as we have shown above, our DMRG results for
the HI-DW transition in the 1D EBHM are consistent with this
expectation. Note that the string order parameter $O_{string}$ is
nonzero in both the HI and DW phases; thus, it is
not the order parameter that is required for identifying
the universality class of the HI-DW transition. We find, furthermore,
the string-order-parameter correlation length is so large
in the HI phase that, given the values of $L$ in our study, it
is not possible to get a reliable estimate for this correlation length.

To investigate the critical behavior of the MI-HI transition, we
plot  $R_{string}(k=0)L^{2\beta/\nu}$ versus $V$, in
Fig.~\ref{fig:OPS} $(b_1)$, whose inset shows a plot of
$R_{string}(k=0)L^{2\beta/\nu}$ versus $(V-{V_c})^\nu L$. These
scaling plots indicate that, for $U=6$, the MI-HI critical point
is at $V_c \simeq 3.5$ with exponents $\beta=0.35$ and
$\nu=1.18$; this value of $\beta$ is consistent with the
Gaussian-model~\cite{gaussian1} (superscript $G$) result
$\beta^G=1/\sqrt{8}$. Our value for $\nu$ follows from the plot
of $L/\xi_L$ versus $V$ in  Fig.~\ref{fig:OPS} $(b_2)$, whose
inset shows a plot of $L/\xi_l$ versus $(V-{V_c}) L^{1/\nu}$. For
the central charge we obtain $c=1.02$, which is consistent with
$c^G=1$ given our error bars, from the plot of $S_L(l)$ versus
$\lambda$ for $L = 220$ in  Fig.~\ref{fig:OPS} $(b_3)$.  As we
have mentioned above, for large $U$ we can restrict the
occupancies in the 1D EBHM to $0,\, 1,$ and $2$ bosons per site to
obtain a spin-1 model. The analog of the MI-HI transition in this
spin-1 model has been shown to be in the Gaussian universality
class~\cite{gaussian1}, which has a fixed exponent $\beta =
1/\sqrt{8}$ and central charge $c=1$, but an exponent $\nu$ that
changes continuously along the phase boundary.  We find indeed,
from calculations like those presented in Figs.~\ref{fig:OPS}
$(b_{1})-(b_{3})$, that $\beta$ and $c$ do not change as we move
along our MI-HI phase boundary in Fig.~\ref{fig:PhaseD} (c), but
$\nu$ does; we obtain, in particular, $\nu=1.18,\, 1.22$, and
$1.36$ for $U=6,\, 5$, and $4$, respectively.

The critical behavior of the SF-HI transition follows from the
plot of $R_{string}(k=0)L^{2\beta/\nu}$ versus $V$, in
Fig.~\ref{fig:OPS} $(c_1)$, and the the plot of $L/\xi_L$ versus
$V$, in  Fig.~\ref{fig:OPS} $(c_2)$, whose inset shows a plot of
$F$ versus $D$ (see Eq.~\ref{eq:measure}). These scaling plots
indicate that, for $V=0.9U$, the SF-HI critical point is at $V_c
\simeq 1.7$, where the correlation length diverges with an
essential singularity as it does at a BKT
transition~\cite{kogutreview}. Thus, we cannot define the
exponent $\nu$; however, we can define the exponent ratio
$\beta/\nu$, which governs how rapidly the string order parameter
vanishes as we approach the SF-HI phase boundary from the HI
phase; for this ratio we obtain the estimate $\beta/\nu \simeq
0.9$. We find the central charge $c=0.94$, which is consistent
with $c=1$ given our error bars, from the plot of $S_L(l)$ versus
$\lambda$ for $L = 280$ in Fig.~\ref{fig:OPS} $(c_3)$. The field
theory for this SF-HI transition is likely to be non-standard
because the SF phase has algebraic $U(1)$ order, whereas the
HI phase has a non-local, string order parameter. An example
of a similar transition is found in the phase diagram of the
bilinear-biquadratic spin-1 chain, in which the Haldane phase
(the analog of the HI phase here) undergoes a transition to a
critical phase, as we change the ratio of the bilinear and
biquadratic couplings; this transition is known to be described
by an $SU(3)_1$ Wess-Zumino-Witten theory~\cite{itoi}.  The
gapless phase in our model is quite different from that in the
spin-1 chain; and there is no $SU(3)$ symmetry at the SF-HI
transition in the 1D EBHM. Nevertheless, it is likely that the
SF-HI phase transition in our model is of a non-standard type,
which is similar to, but not the same as, the $SU(3)_1$ theory
mentioned above; this point requires more detailed investigations
that lie beyond the scope of our paper.  Note, furthermore, that
the SF-HI transition here is not fine-tuned, unlike the one above
in the spin-1 chain with bilinear and biquadratic couplings,
in so far as the SF-HI transition occurs along an entire boundary
in the phase diagram (Fig.~\ref{fig:PhaseD} (c)).

For the sake of completeness we also show, for the BC3 boundary
condition, the BKT nature of the SF-MI transition in
Fig.~\ref{fig:PhaseD} (c), as noted for the BC1 case in
Ref.~\cite{rvpai05}, by presenting plots of the scaled charge gap
$F$ versus $D$ (see Eq.~\ref{eq:measure}), the $k=0$ value of
$n(k=0)$, and the block entanglement entropy $S_L(l)$ versus the
log-conformal distance $\lambda$ in Figs.~\ref{fig:OPS} $(d_1)$,
$(d_2)$, and $(d_3)$, respectively. These plots demonstrate that,
at the MI-SF transition, the correlation length (or inverse
charge gap) has a BKT-type essential singularity
(Figs.~\ref{fig:OPS} d$_1$), the SF-correlation-function exponent
$\eta=1/4$, and the central charge is $c=0.96$, which is
consistent with $c^{BKT}=1$ given our error bars. It is difficult
to give precise error bars for the exponents and central charges
that we have calculated because there are systematic errors,
which include those associated with the finite values of
$n_{max}$ and $n_{states}$; these errors are hard to estimate.

At lower values of $U$ and $V$ than those shown in
Fig.~\ref{fig:PhaseD}, it has been suggested that the 1D EBHM can
show a supersolid (SS) phase between SF and DW phases (see, e.g.,
Ref.~\cite{deng12}). However, it is not easy to distinguish such
an SS phase from a region of first-order phase coexistence
between SF and DW phases in a DMRG calculation, with a fixed
number of bosons; we discuss this elsewhere~\cite{tobepublished}.

We have performed the most extensive DMRG study of the 1D EBHM
attempted so far. Our work, which extends earlier
studies~\cite{rvpai05,entspect,torre06,berg08} significantly, has
been designed to investigate the effects of boundary conditions
on the phase diagram of this model, to elucidate the natures of
the phase transitions here, and to identify the universality
classes of the continuous transitions. Our study shows that the
HI-DW, MI-HI, and SF-MI transitions are, respectively, in the 2D
Ising, Gaussian, and BKT universality classes; however, the SF-HI
transition seems to be more exotic than a BKT transition in so
far as the string order parameter also vanishes where the SF
phase begins. We find that the different boundary conditions BC1,
BC2, and BC3, which lead to the same value of $\rho$ in the thermodynamic limit, yield different phase diagrams; this
boundary-condition dependence deserves more attention than it has
received so far, especially from the point of view of the
existence of thermodynamic limits~\cite{thermodynamiclimit}. We
hope our work will stimulate more theoretical and experimental
studies of the 1D EBHM and experimental realizations thereof.

We thank DST, UGC, CSIR (India) for support and E. Berg, E. G. Dalla Torre,
F. Pollmann, E. Altman, A. Turner, and T. Mishra for useful discussions.

%%%%%%%%%


\begin{thebibliography}{99}
\bibitem{footnote} We use the conventional terminology charge gap
even though we do not have charged bosons in mind. Indeed, some
authors refer to our DW phase as a CDW phase.
\end{thebibliography}
\end{document}